\documentclass[reprint,prl,aps,epsf,twocolumn,superscriptaddress]{revtex4-1}

\usepackage{graphicx}
\usepackage{amssymb}
\usepackage{amsmath}

\begin{document}
\title{Invasion-wave induced first-order phase transition in systems of active particles}
\author{Thomas Ihle}
\affiliation{Department of Physics, North Dakota State University,  
Fargo, North Dakota, 58108-6050, USA}    

\begin{abstract}
An Enskog-like kinetic equation for self-propelled particles 
is solved numerically.
I study a density instability near the transition to collective motion and find that while hydrodynamics 
breaks down, the kinetic approach leads to soliton-like supersonic waves with steep leading kinks  
and Knudsen numbers
of order one.  
These waves show hysteresis, 
modify the transition threshold and lead to an abrupt jump of the global order parameter if the noise level is changed. 
Thus they provide a mechanism to change the second-order character of the phase transition to first order.
\end{abstract}

\maketitle

Collective motion of self-propelled 
agents is a key feature
of  
active matter systems and has attracted much 
attention \cite{ramaswamy_10,vicsek_zafeiris_12,marchetti_12}. 
Systems of interest range from 
animal flocks \cite{animal_flocks}, to 
human crowds \cite{human_crowds},  
actin networks driven by molecular motors \cite{actin_net}, 
interacting robots \cite{robot_swarms}, and mixtures of robots and living species \cite{animal_robot}. 

Most of our theoretical understanding of collective motion comes from two sources: 
(i) computational studies of particle-based models 
and
(ii) phenomenological transport equations which 
are usually postulated by means of symmetry 
arguments as in the seminal work by Toner and Tu \cite{toner_95_98,toner_12}. 
These authors showed that
even in a two-dimensional system, 
long-range orientational order is possible 
due to the nonzero speed of the particles.
Many of the computational approaches \cite{chate_04_08,chate_08,baglietto_08_09,ginelli_10,peruani_11,levine_00,dossetti_09,chate_10,lobaskin_13,meschede_12} are related to the minimal Vicsek-model (VM) 
\cite{vicsek_95_07}. In the VM, pointlike particles move at fixed speed and 
try to align locally with their neighbors
but do not succeed completely due to the presence of some noise.
As the noise amplitude decreases, the system experiences a phase transition from a 
disordered state, in which the particles have no prefered global direction, to an
ordered state, in which, on average,
the particles move 
in the same direction. 

This paper is based on a third angle of attack -- the kinetic theory of gases.  
Apart from a few exceptions \cite{helbing_96,bussemaker_97,bertin_06}   
the kinetic approach to active particle systems
has  been less popular. However, it
is very powerful and 
allowed the solution of a long standing problem, the rigorous derivation of
the hydrodynamic theory for the VM \cite{ihle_11}.
See also Refs. \cite{mishra_10,romanczuk_12,grossmann_13} for alternative derivations.

Direct simulations of the VM \cite{chate_04_08,chate_08,aldana_09} revealed that right at the onset of ordered motion, large density waves emerge.
It has been intensely debated \cite{chate_04_08,chate_08,vicsek_95_07,baglietto_08_09,aldana_09}
whether this order-disorder transition is continuous or discontinuous.
By now it is generally accepted that at high particle speeds 
the transition
is discontinuous with strong finite size effects.
This is in puzzling contrast
to mean-field theories \cite{bertin_06,bertin_09,ihle_11,pimentel_08,aldana_03} which should be valid at large speeds but predict a continuous transition.

In this Letter, 
I show that the solution of this puzzle lies beyond hydrodynamic theory. 
I find that a special soliton-like density wave, which can be analyzed by kinetic theory but not by hydrodynamics, 
is able to alter the character of the phase transition from continuous to discontinuous.
I calculate the global order parameter for collective motion and show 
explicitly how its hysteresis and its unique finite size effects
are related to the properties of the density waves.

The reason why these waves escape hydrodynamic treatment 
is that they violate a  
basic principle for the validity of a hydrodynamic theory -- the smallness of the Knudsen number -- 
which requires that  
the average distance 
particles travel between collisions is
much smaller than the length over which hydrodynamic fields 
change considerably. 
This is not true
for the waves emerging near the onset of collective motion.
They are so steep that their Knudsen number is always of order one.

In one of the first analytical studies 
of active particles, Bertin {\em et al.} \cite{bertin_06,bertin_09}
have also analyzed soliton waves. 
However, the calculated density profiles (Fig. 7 of Ref. \cite{bertin_09}) bear 
little resemblence with the actual profiles obtained in direct simulations of the VM, which have a very sharp leading edge.
Gopinath {\em et al.} \cite{gopinath_12} 
also calculated waves which look
different from the ones observed in simulations.
Both groups obtained waves {\em within} the hydrodynamic approach and did not
observe that the waves have any effect on the order-disorder threshold.
The waves calculated in this Letter by means of kinetic theory are qualitatively different from the ones of Refs. \cite{bertin_09,gopinath_12}
because (i) they shift the transition threshold and modify the character of the 
flocking transition from second to first-order, and (ii) their profile semi-quantitatively agrees with the ones 
measured in direct simulations \cite{peak_pic}.
 
A first clue about the inadequacy of hydrodynamic equations for the VM 
comes from
Ref. \cite{ihle_11} where it was shown that if 
all coefficient in these equations are
rigorously derived from the microscopic dynamics,
long wavelength density modulations evolve 
into waves with infinite amplitudes.
Thus, contrary to Refs. \cite{bertin_09,gopinath_12} no solitons
could be found.
The equations were derived 
under the assumption that higher order gradient terms are negligible
which is not justified when steep spatial perturbations of a homogeneous state grow sufficiently large.
Usually, such perturbations are stabilized
by higher order nonlinear terms.
This is not the case here,
the hydrodynamic equations are driven out of their range of validity.
To discover the final fate of these waves
within the hydrodynamic approach one would have to 
explicitly sum gradient terms of all orders,
which is practically impossible. 
I circumvent this obstacle by
abandoning
hydrodynamics
entirely. 
Instead, I numerically solve the space and time-dependent kinetic equations of the VM. 
Because this does not involve any gradient expansion, a summation to all orders is achieved implicitly.

In the VM, $N$ 
pointlike particles with 
positions ${\bf x}_i(t)$ and velocities ${\bf v}_i(t)$ 
undergo a discrete-time dynamics
with time step $\tau$. The evolution consists of two steps: 
streaming, where all positions are updated according to 
${\bf x}_i(t+\tau)={\bf x}_i(t)+\tau {\bf v}_i(t)$,
and collision.
The magnitude $v_0$ of the particle velocities is kept constant,
only the directions $\theta_i$ of
the velocity vectors
are modified in the collision step:
a circle of radius $R$ is drawn around a given particle and the average direction $\Phi_i$ of motion
of the
particles
within the circle is determined
according to
$\Phi_i={\rm arctan}[\sum_j^n {\rm sin}(\theta_j)/\sum_j^n {\rm cos}(\theta_j)]$.
The new directions
follow as
$\theta_i(t+\tau)=\Phi_i(t)+\xi_i$. Here, $\xi_i$ is a random number which is
uniformly distributed in 
the interval $[-\eta/2,\eta/2]$.

Following Ref. \cite{ihle_11}, the time evolution of the VM can be described by 
a Markov chain
for the N-particle probability density.
This equation is exact but
intractable without simplification.
The easiest way to proceed is to make Boltzmann's molecular chaos approximation
and assume that the particles are uncorrelated 
{\it prior} to a collision,
which amounts to a factorization of the N-particle probability into a product of one-particle probabilities.
Because this assumption neglects
correlations and leads to an effective one-particle picture,
it can be thought of as a sort of mean-field theory which
looks like an Enskog equation,
\begin{eqnarray}
\nonumber
& &
f(\theta, {\bf x},t+\tau)=
{1\over \eta}
\int_{-\eta/2}^{\eta/2}
d\xi
\sum_{n=1}^N
{1\over (n-1)!}{\rm e}^{-M_R({\bf x'},t)}
\\ \nonumber
& &\times \int_{0}^{2\pi}d\tilde{\theta}_1
 \bigg [ \prod_{i=2}^n\int_{0}^{2\pi}d\tilde{\theta}_i\int_{\bigodot}d\tilde{\bf x}_i
\,f( \tilde{\theta}_i, {\bf \tilde{x}}_i,t) \bigg ] \\
\label{ENSKOG1}
& &\times f(\tilde{\theta}_1,{\bf x'}, t)
\,\,\hat{\delta}(\theta-\xi-\Phi_1(\tilde{\theta}_1,\ldots \tilde{\theta}_n))\,.
\end{eqnarray}
The distribution function $f(\theta,{\bf x})$ is
proportional to the probability 
to find a particle with a given angle $\theta$ at location ${\bf x}$.
Details on this derivation
can be found in Refs. \cite{ihle_11,chou_12,ihle_09}.
The r.h.s. of Eq. (\ref{ENSKOG1}) is the collision integral and will be denoted as $I[f]$.
It is a nonlinear functional of the distribution function with a singular kernel
which consists of 
the periodically continued Dirac-delta function,
$\hat{\delta}(\alpha)=\sum_{m=-\infty}^{\infty}\delta(\alpha+2\pi m)$.
The argument of the exponential in Eq. (\ref{ENSKOG1}),
$M_R({\bf x}',t)=\int_{\odot}\rho({\bf y},t)\,d{\bf y}$,
is the average number of particles in a circle of radius $R$ centered around 
${\bf x}'={\bf x}-{\bf v}\tau$ where ${\bf v}=v_0(\cos\theta,\sin{\theta})$
is a velocity vector.
This interaction circle is not centered around the final position ${\bf x}$
because after the reference particle $i=1$ has collided with particles $2,3\ldots n$
it will be convected
to location ${\bf x}$ in the subsequent streaming step.
The symbol $\odot$ denotes spatial integrations over the collision circle.
The particle density $\rho$ is given as the zeroth moment of the distribution function,
$\rho({\bf x},t)=\int_0^{2\pi} f(\theta,{\bf x},t)\,d\theta$. 

Eq. (\ref{ENSKOG1}) is solved by an algorithm which is related to 
the Lattice-Boltzmann method (LB) \cite{lattice_boltz}. It relies on a set of $Q$ microscopic velocities, 
${\bf e}_i$, 
where every velocity is associated with a distribution function $f_i({\bf x},t)$.
The positions ${\bf x}$ are discretized on a regular lattice of size $L_x\times L_y$. 
See Supplemental Material at [URL] for details on this numerical method.
In contrast to LB a very large number
of velocities, $Q\approx 1000$,
is used to accurately resolve the order-disorder transition.
Another difference is the nonlocal collision term $I[f]$ which requires spatial and angular integrations. 
Naive attempts to calculate $I[f]$
by a simple integration scheme lead to prohibitively slow performance.
A much more accuate and faster way is
to evaluate the collision integral in angular Fourier space.
Expanding $f$,
\begin{equation}
\label{FOURIER_MODE_F}
f(\theta,{\bf x})=
\sum_{k=0}^{k_C}\left[ g_k({\bf x})\,{\rm cos}(k\theta)+h_k({\bf x})\,{\rm sin}(k\theta)\right]\,,
\end{equation} \\
performing a similar expansion for $I[f]$, 
and evaluating Eq. (\ref{ENSKOG1}) in this basis gives
a simple set of algebraic relations for the Fourier coefficients of the collision integral in terms
of $g_k$ and $h_k$, see Eq. (S2) in the Supplemental Material at [URL].
Three-particle and higher order collisions have been neglected.
While this restriction to binary interactions 
keeps 
the simulation times short and
reduces the validity of the numerics 
to low densities, $M_R<1$, 
it is not a principal limitation.
Similar to Ref. \cite{chou_12}, the algorithm can be easily extended to include genuine three-, and higher particle 
collisions.
In Eq. (\ref{FOURIER_MODE_F}) all angular modes with wavenumbers $k$ larger than the cut-off value $k_C=5$
were neglected. 
The remaining modes are sufficient to describe the behavior of the order parameter in the vicinity of a 
phase transition.
The 
global order parameter $\Omega$ is defined by means of the $k=1$ Fourier coefficients, $\Omega =\langle \sqrt{g_1^2+h_1^2} \rangle$
where the brackets denote an average over the simulation box.
These coefficients
are proportional to the components of the momentum density ${\bf w}$,
$g_1\propto w_x$, $h_1\propto w_y$, which is 
given by the first moment of the distribution function, 
${\bf w}({\bf x},t)=\int_0^{2\pi} {\bf v}(\theta)f(\theta,{\bf x},t)\,d\theta$. 
\begin{figure}
\begin{center}
\vspace{0cm}
\includegraphics[width=1.6in,angle=0]{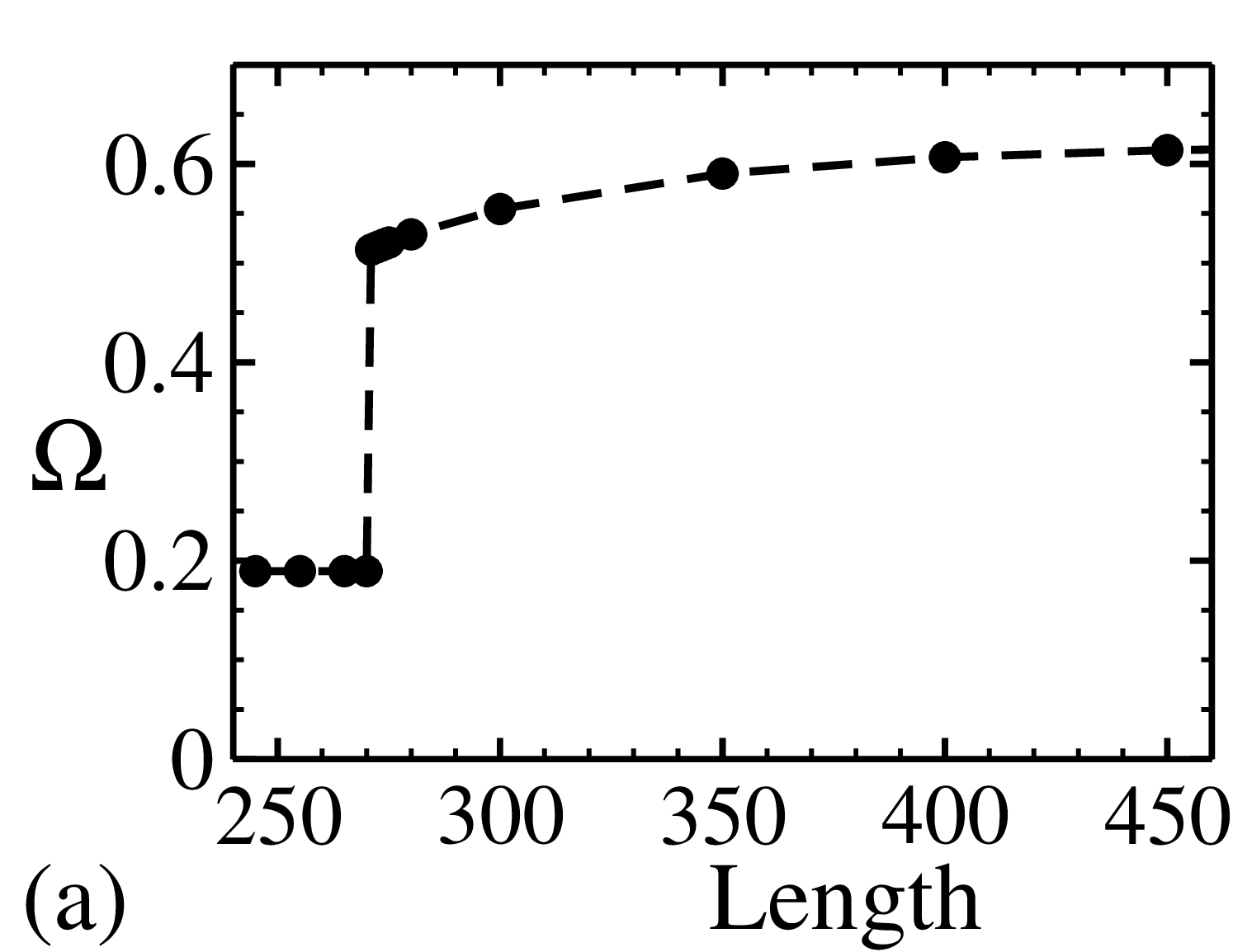}
\hspace{1ex}
\includegraphics[width=1.6in,angle=0]{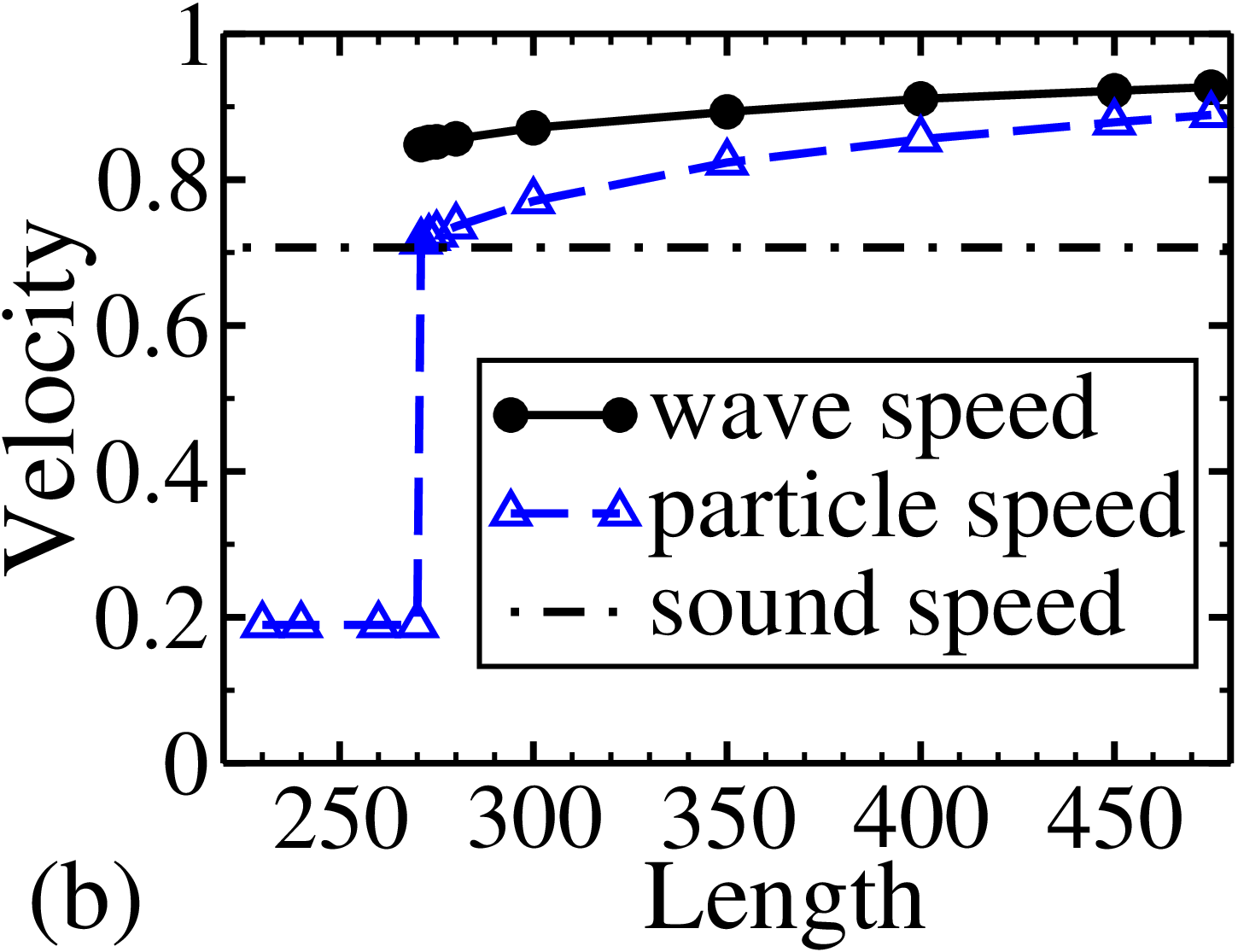}
\vspace{-1ex}
\caption{(a) Steady state order parameter $\Omega$; (b) speed of the invasion wave $v_W$ (circles) and 
average particle speed $u=|{\bf w}|/\rho$ (triangles) measured at the tip of the wave
versus system size. 
The dash-dotted line shows the speed of sound in the disordered phase, $v_S=v_0/\sqrt{2}$.
All speeds are plotted in units of $v_0$.
Parameters: $R=1.5$, $v_0=0.97$, $\eta=0.43$.
}
\label{FIG1}
\end{center}
\vspace*{-5ex}
\end{figure}
\begin{figure}
\begin{center}
\vspace{0cm}
\includegraphics[width=1.6in,angle=0]{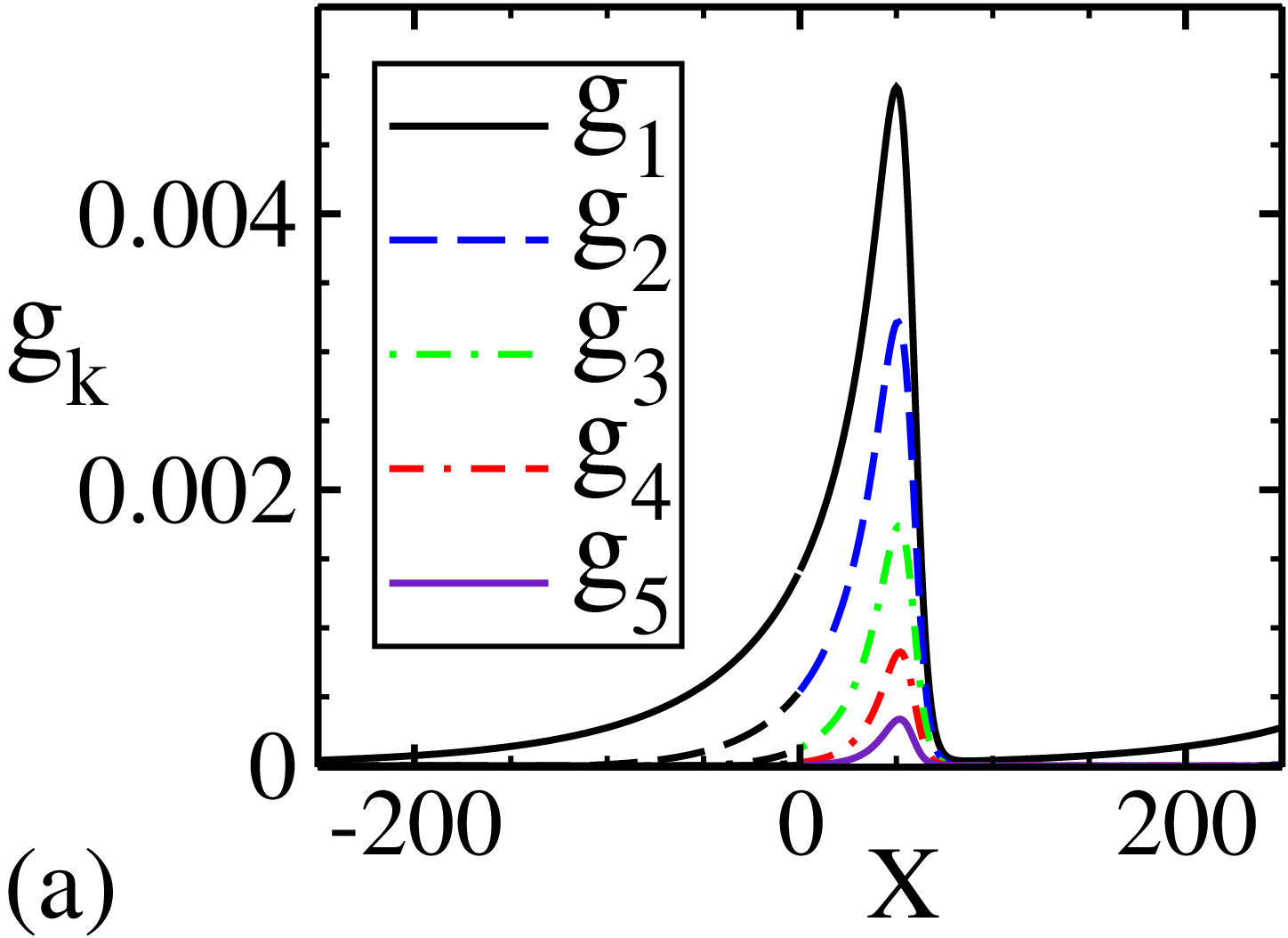}
\hspace{1ex}
\includegraphics[width=1.6in,angle=0]{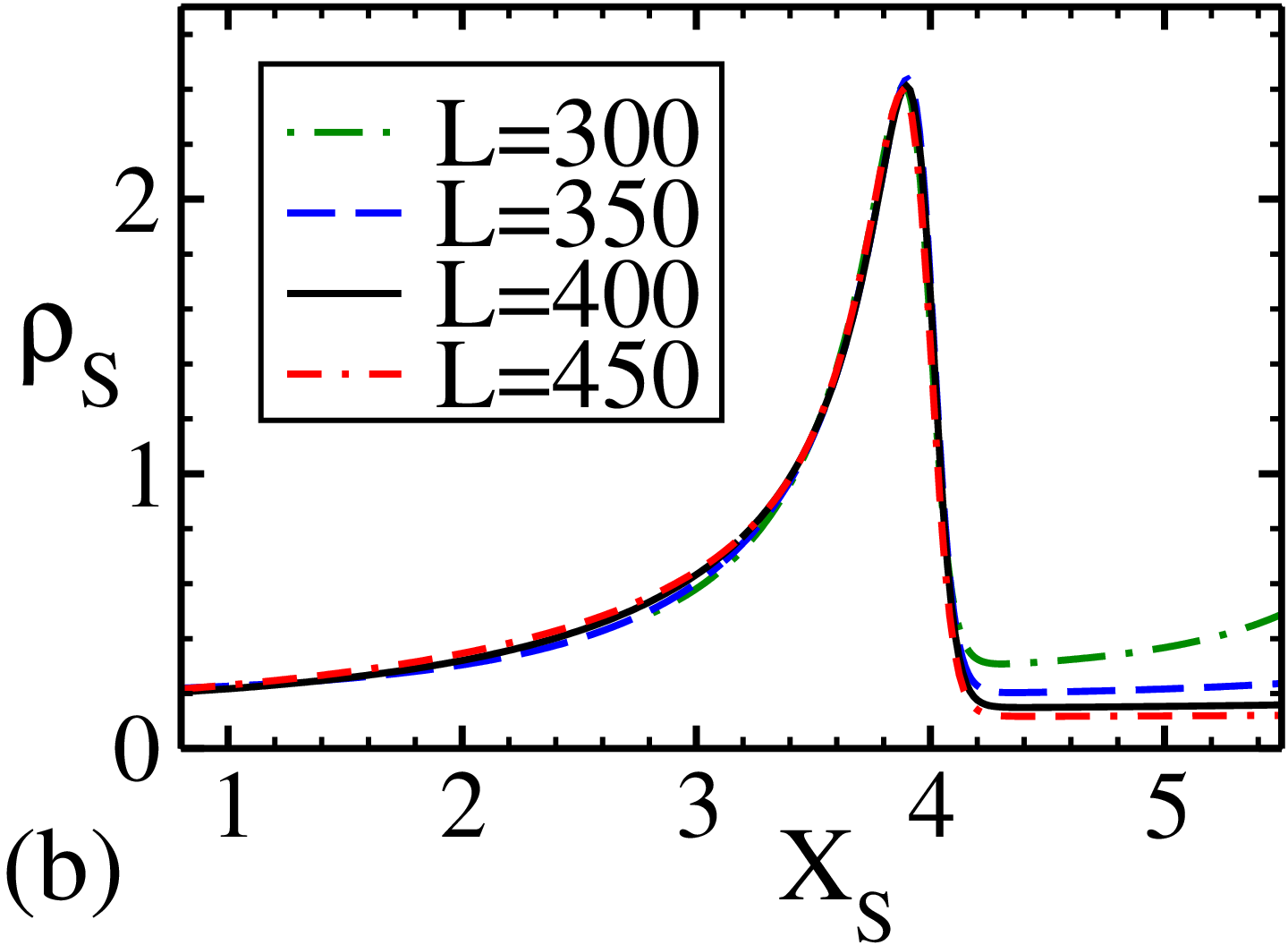}
\vspace{-1ex}
\caption{(a) The Fourier coefficients $g_k$ for a stationary invasion wave travelling
into the positive x-direction as a function of position.
(b) The rescaled density $\rho_S=\rho L_x^{-2} 10^6$ versus rescaled position $X_S=x L_x^{2} 10^{-5}$
for different system sizes. Parameters are the same as in Fig. \ref{FIG1}.
}
\label{FIG2}
\end{center}
\vspace*{-5ex}
\end{figure}
Let us first consider a very small system, $L_x=L_y=4$ with periodic boundary conditions and
average particle density $\rho_0=N/(L_x L_y)=0.00424$.  
To initialize a disordered state, all angular Fourier coefficients
except $g_0=\rho_0/2\pi$ are set to zero. 
In this and all following simulations, the average particle number in the collision circle, $M=\pi R^2\rho_0$, 
is set
to $M=0.03$ and the time step is $\tau=1$.
Recently \cite{ihle_11}, the critical noise $\eta_C$ as a function of $M$ has been calculated.
By choosing a noise value $\eta=0.43$ slightly smaller
than the threshold value $\eta_C=0.4361$, the system is quenched into the ordered state. 
Since the system size is much smaller than the critical length $L_0=2\pi/k_0$ (see Fig. 2 of 
\cite{ihle_11}) above which
the homogeneous ordered state is linearly unstable, the system is expected to stay homogeneous.
The numerical solution of Eq. (\ref{ENSKOG1}) agrees with these predictions: 
the zero momentum disordered state
evolves into a stable ordered state.
As shown in Fig. \ref{FIG1}(a), the  
order parameter stays constant at increasing system size until a critical size
of $L^*\approx 270.5$ is reached.
When the system size is adjusted by just one lattice unit from $L=270$ to $271$, 
$\Omega$ makes an abrupt jump, almost tripling its value.
A closer look reveals that while the steady state solution is homogeneous at small $L$, it is
not homogeneous above $L^*$. 
Instead, a single-peaked density wave is going through the system.
It travels at constant velocity $v_W$ and, as seen in Fig. \ref{FIG2}, has a 
pronounced assymetric shape with a steep front and an extended 
tail.
The anharmonicity of this shape, together with previous results \cite{ihle_11}, provide
a simple explanation for the disconinuity of the global order parameter: 
The wave is born as a result of a linear instability and inherent noise. Once it exists, it grows to a large final size 
because nonlinear attenuation is ineffective.
The definition of $\Omega$ involves a spatial average which is dominated by the extended spatial 
region behind 
the peak of the wave, where local order is much stronger than in the corresponding homogeneous ordered state.
Even though the area ahead of the wave front is mostly disordered with $|{\bf w}|\approx 0$, 
it cannot compensate
the huge contributions to $\Omega$ from the densest part of the wave. 
As a result of this biased average, the global order parameter is much larger
than in the homogeneous ordered state.
\begin{figure}
\begin{center}
\vspace{1ex}
\includegraphics[width=3.0in,angle=0]{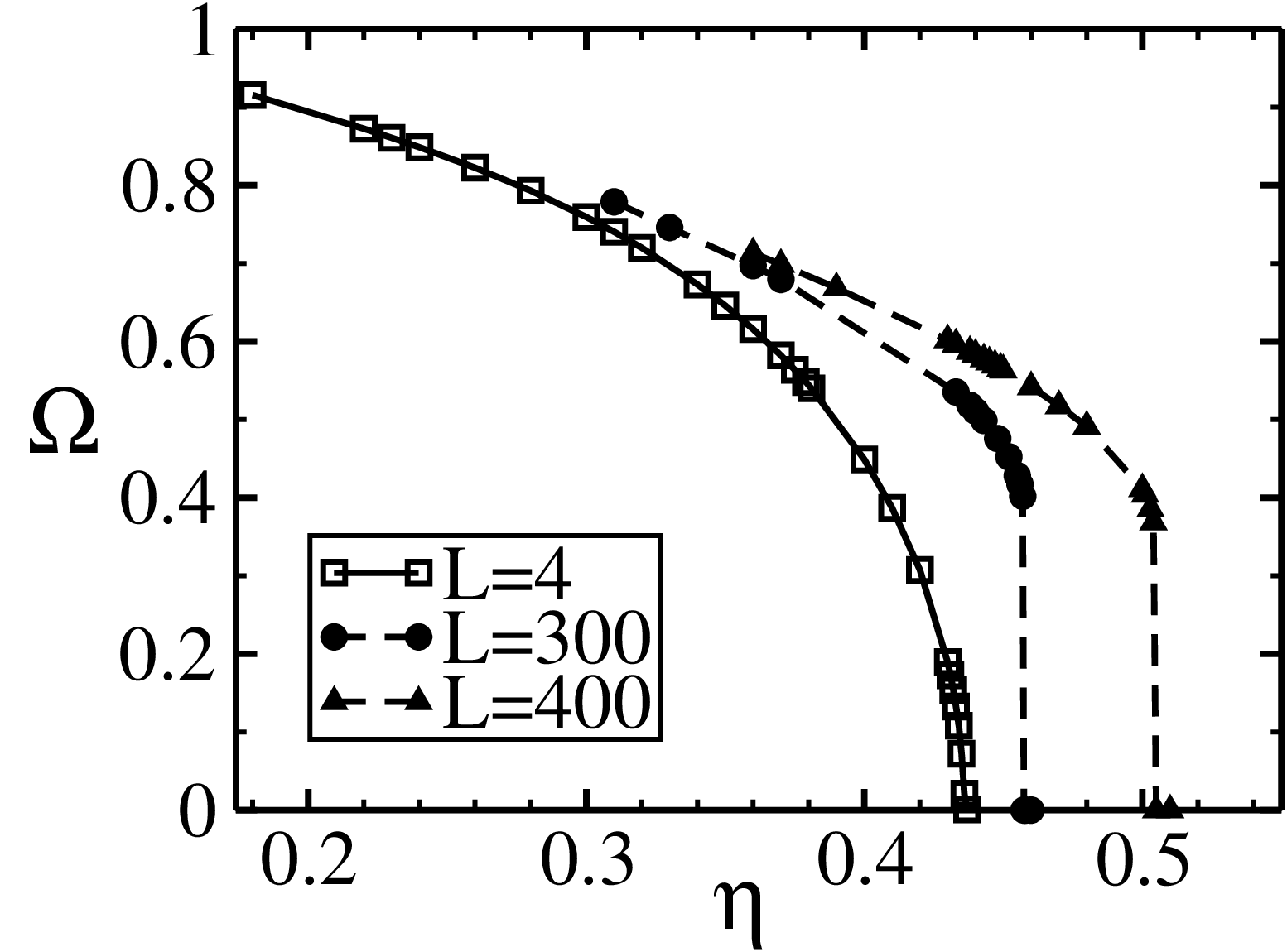}
\vspace{-1ex}
\caption{Order parameter versus noise $\eta$ for several system sizes $L$.
Parameters: $R=0.1$, $v_0=1$.
}
\label{FIG3}
\end{center}
\vspace*{-5ex}
\end{figure}
These waves have remarkable properties. 
For example, Fig. \ref{FIG1}(b) shows that their speed $v_W$ is just slightly below the maximum possible speed $v_0$,
but larger than the speed of sound in the disordered phase, $v_S=v_0/\sqrt{2}$.
The waves are always supersonic with Mach numbers
between $1.19$ and $1.41$.
The particles at the highest (that is densest) point of the wave are so strongly aligned that their average speed 
$u=|{\bf w}|/\rho$ is only slightly less than $v_W$.
Thus, most particles in the wave crest travel with the wave and do not just undergo restricted local motion.
The particles just ahead of the wave front have low density and display disordered motion.
They do not ``feel'' the wave coming since it arrives faster than the speed of sound. Their territory is invaded and a 
fraction of them becomes strongly aligned and joins the wave for a while. 
This motivates the term {\em invasion wave}.

In agreement with direct simulations \cite{wave_num}, 
I observe that the invasion wave has a perfectly straight front, perpendicular to its direction of motion.
To accelerate the numerics, 
I take advantage of this apparent one-dimensional nature and  
drastically reduce the y-extension of the box
to $L_y=2$, creating a very elongated simulation box.
Examining large  box lengths $L_x$, one realizes 
that the maximum density in a wave and the steepness of the leading edge depends on system size
in a very sensible way that transcends the traditional meaning of ``finite size effects''.
In particular, the maximum density in the wave is proportional to $L_x^2$, and the width of the peak
scales as $1/L_x^2$.
In fact, the invasion wave cannot be seen as a localized perturbation of some mainly undisturbed medium.
It is rather a global excitation of the entire system where, facilitated by periodic boundary conditions, 
all parts of the system are involved and particles everywhere
adjust accordingly.
In Fig \ref{FIG2}(b), by scaling the x-coordinate by $L_x^{-2}$ and the density by $L_x^2$ 
it is demonstrated
that here is a master curve for the shape
of an invasion wave. 
\begin{figure}
\begin{center}
\vspace{0cm}
\includegraphics[width=2.6in,angle=0]{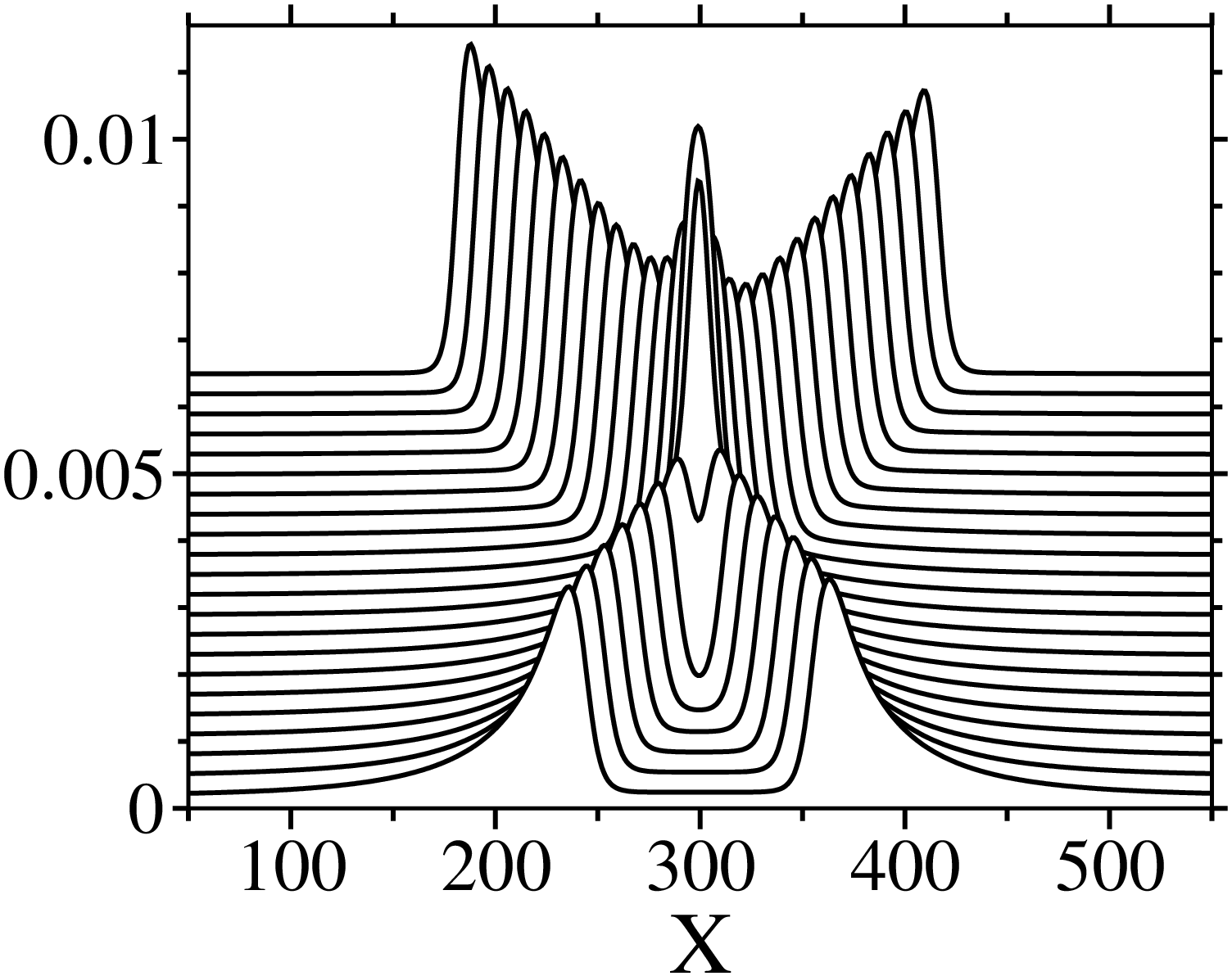}
\vspace{-1ex}
\caption{
Snapshots for the head-on collision of two soliton-like waves. 
The sequence starts with two well seperated peaks close to the x-axis running
towards collision with their steep fronts facing each other.
At the latest time, the peaks are separated again after a successful ``tunneling'' through each other and now run towards
the edges of the box.
\vspace*{-3ex}
}
\label{FIG4}
\end{center}
\end{figure}
As a consequence,
the maximum density gradient at the leading edge is proportional to $L_x^4$.
To understand why these waves have not been detected by means of hydrodynamic approaches \cite{bertin_09,gopinath_12} 
an effective
Knudsen number, $Kn=\lambda/r$ is defined. Here, $\lambda=v_0\tau$ is the mean free path of a particle and $r$ is the radius of curvature
of the density profile at the 
tip of the wave.
It turns out that $Kn$ is never smaller than about $1/2$, not even in the smallest systems that allow wave formation. 
At such Knudsen numbers, a structure with a small internal scale moves quite a large distance
in one time step, causing hydrodynamic gradient expansions to diverge.

Assuming a homogeneous ordered state, the mean-field theory of the VM \cite{ihle_11} 
predicts that the flocking transition is continuous.
To check whether this is still true for inhomogeneous ground states,
I calculate the order parameter as a function of noise for different system sizes.
According to Fig. \ref{FIG3}, at the smallest size $L=4$ where the system is still homogeneous, a continuous transition occurs.
However, 
in bigger systems, very close to the predicted threshold $\eta_C=0.4361$, $\Omega$ still has a large value 
because of persistent waves.
To accurately establish the order of this transition I used the following simulation protocol:
An ordered state with a stable stationary wave at very low noise $\eta_0$ is created.
This inhomogeneous state serves as initial condition for a run with slightly larger noise $\eta_1>\eta_0$.
After convergence, the noise is increased again. This way, a sequence of stable
inhomogeneous states with large order parameters is obtained, even at noise values a few percent above $\eta_C$.
Approaching the threshold from the higher noise side confirms that the disordered state with $\Omega=0$ is stable
for $\eta\ge \eta_C$. Hence, the flocking transition shows hysteresis; the wave state can 
coexist with the homogeneous state over a finite noise range $\Delta \eta\approx 0.045 \eta_C$ for $L=300$.
This concludes the
proof for the discontinuity of
the order-disorder 
transition 
for large $L$, at the mean-field level.
As seen in Fig. \ref{FIG3} the hysteresis region grows with system size 
because the properties of the underlying soliton-like wave strongly depends on $L$. 
Compared to equilibrium systems, the phase behavior depicted in Fig. \ref{FIG3} looks unusual.
Nevertheless, it semi-quantitatively
agrees with direct simulations of the VM \cite{phasediag_num,bag_13}. 

One of the defining properties of a soliton is the ability to pass through each other without destruction.
To perform this ``soliton test'', I prepared stationary waves in two different systems with slightly different
sizes, $L_x=299$ and $L_x=300$ and ensured they run in opposite directions.
After the waves became stationary, I ``glued'' the two boxes together leading to a longer 
system with $L_x=599$. A series of snapshots of the time evolution of this two peak system is shown in
Fig. \ref{FIG4}.
At the earliest time one sees two peaks running towards each other. Eventually, 
they start to overlap and form a large single peak.
A while later, the two peaks reemerge with almost undisturbed shape like a conventional soliton.
Watching the time evolution through repeated collisions reveals that if the waves have a tiny height difference initially,
this difference is amplified in every encounter.
The bigger soliton takes a few particles
away 
from the smaller one in every meeting until only one peak survives.
Gradual coarsening also occurs inbetween collisions or when waves travel in the same direction.
Therefore, true stationary states have only one peak.

In conclusion, using kinetic theory I analyzed supersonic waves 
that
occur in the Vicsek model once the system exceeds a critical size.
I demonstrated that the waves show hysteresis and alter the order of the flocking transition from continuous to discontinuous.
A phase diagram with atypical finite size effects was calculated.
I argue that these waves were probably not detected in previous analytical approaches because of their large Knudsen number.
This provides  
an explicit example of an active particle model
where hydrodynamic equations fail to correctly describe one of 
the most important properties -- the nature of the phase transition.
This could have implications for other, more sophisticated, models of active matter.
While my calculations neglect correlations which are relevant at small particle speeds,
they do demonstrate a novel mechanism to induce a first-order flocking transition.
How relevant this mechanism is in the low speed regime and for particles with nonzero size remains an open question.
I speculate that some aspects of this approach remain useful. 
For example, it might lay the ground for a theory
in terms of interacting 
quasi-particles that represent soliton waves.

Support
from the National Science Foundation under grant No.
DMR-0706017 
is gratefully acknowledged.


\begin{thebibliography}{99} 

\bibitem{ramaswamy_10}
S. Ramaswamy, Annu. Rev. Condens. Matter Phys. 1, 323 (2010).

\bibitem{vicsek_zafeiris_12}
T. Vicsek and A. Zafeiris, Phys. Rep. {\bf 517}, 71 (2012).

\bibitem{marchetti_12}
M.C. Marchetti {\em et al.}, arXiv:1207.2929, (2012).

\bibitem{animal_flocks}
J.K. Parrish and L. Edelstein-Keshet, Science 284, 99
(1999);
I.D. Couzin {\em et al.}, 
Nature {\bf 433}, 513 (2005);
I.D. Couzin, J. Krause,
Adv. Stud. Behav. {\bf 32}, 1 (2003);
R.W. Tegeder, J. Krause, Philos. Trans. R. Soc. London
B 350, 381 (1995);
N. Abaid, M. Porfiri, J. R. Soc. Interface {\bf 7} 1441 (2010). 

\bibitem{human_crowds}
D. Helbing, I. Farkas, T. Vicsek,
Nature {\bf 407}, 487 (2000);
D. Helbing, Rev. Mod. Phys. 73, 1067 (2001). 

\bibitem{actin_net}
V. Schaller {\em et al.},
Nature {\bf 467}, 73 (2010);
V. Schaller, C. Weber, E. Frey and A.R. Bausch,
Soft Matt. {\bf 7}, 3213 (2011);
J. F. Joanny {\em et al.},
New J. Phys. {\bf 9} 422 (2007).

\bibitem{robot_swarms}
A. Jadbabaie, J. Lin, S. Morse, IEEE Trans. Auto. Control {\bf 48},
988 (2003); A.E. Turgut {\em et al.}, Swarm Intell. {\bf 2}, 97 (2008); W.M. Shen {\em et al.} Auton. Robots {\bf 17}, 93 (2004).

\bibitem{animal_robot}
R. T. Vaughan {\em et al.},
Robot. Auton. Syst. {\bf 31}, 109 (2000);
J. Halloy {\em et al.},
Science {\bf 318}, 1155 (2007);
S. Marras, M. Porfiri,
J. R. Soc. Interface {\bf 9}, 1856 (2012).

\bibitem{toner_95_98}
J. Toner and Y. Tu,
Phys. Rev. Lett. {\bf 75}, 4326 (1995); 
Phys. Rev. E
{\bf 58}, 4828 (1998).

\bibitem{toner_12}
J. Toner, Phys. Rev. E {\bf 86}, 031918 (2012).

\bibitem{chate_04_08}
G. Gr{\'e}goire and H. Chat{\'e}, Phys. Rev. Lett. {\bf 92} 025702 (2004);
Phys. Rev. Lett. {\bf 99}, 229601 (2007);
Eur. Phys. J. B {\bf 64} 451 (2008).

\bibitem{chate_08}
H. Chat{\'e}, F. Ginelli, G. Gr{\'e}goire, F. Raynaud, Phys. Rev. E {\bf 77} 046113 (2008);

\bibitem{baglietto_08_09}
G. Baglietto, E.V. Albano,
Phys. Rev. E {\bf 78}, 021125 (2008); 
Phys. Rev. E {\bf 80}, 050103 (2009). 

\bibitem{ginelli_10}
F. Ginelli {\em et al.},
Phys. Rev. Lett. {\bf 104} 184502 (2010);

\bibitem{peruani_11}
F. Peruani {\em et al.},
J. Phys. Conf. Ser. {\bf 297} 012014 (2011).

\bibitem{levine_00}
H. Levine, W.-J. Rappel, I. Cohen,
Phys. Rev. E {\bf 63}, 017101 (2000). 

\bibitem{dossetti_09}
V. Dossetti, F.J. Sevilla, V.M. Kenkre,
Phys. Rev. E {\bf 79}, 051115 (2009).

\bibitem{chate_10}
F. Ginelli, H. Chat{\'e},
Phys. Rev. Lett. {\bf 105}, 168103 (2010).

\bibitem{lobaskin_13}
V. Lobaskin, M. Romenskyy,
Eur. Phys. J. B {\bf 86}, 91 (2013).

\bibitem{meschede_12}
M. Meschede, O. Hallatschek,
arXiv:1212.2060v1 (2012).

\bibitem{vicsek_95_07}
T. Vicsek {\em et al.},
Phys. Rev. Lett. {\bf 75}, 1226 (1995); 
A. Czir{\'o}k, H. E. Stanley, T. Vicsek, J. Phys. A, {\bf 30}, 1375 (1997);
M. Nagy, I. Daruka, T. Vicsek, Physica A {\bf 373}, 445 (2007).

\bibitem{helbing_96}
D. Helbing,
Phys. Rev. E {\bf 53}, 2366–2381 (1996). 

\bibitem{bussemaker_97}
H.J. Bussemaker, A. Deutsch, and E. Geigant,
Phys. Rev. Lett. {\bf 78}, 5018 (1997).

\bibitem{bertin_06}
E. Bertin, M. Droz, and G. Gr{\'e}goire, Phys. Rev. E {\bf 74}, 022101 (2006).

\bibitem{bertin_09}
E. Bertin, M. Droz, and G. Gr{\'e}goire, 
J. Phys. A {\bf 42}, 445001 (2009).

\bibitem{ihle_11}
T. Ihle, Phys. Rev. E {\bf 83}, 030901 (2011).

\bibitem{pimentel_08}
J.A. Pimentel {\em et al.},
Phys. Rev. E {\bf 77} 061138 (2008).

\bibitem{aldana_03}
M. Aldana, C. Huepe, J. Stat. Phys. {\bf 112}, 135 (2003).

\bibitem{aldana_07}
M. Aldana {\em et al.},
Phys. Rev. Lett. {\bf 98}, 095702 (2007).

\bibitem{gopinath_12}
A. Gopinath {\em et al.},
Phys. Rev. E {\bf 85}, 061903 (2012).

\bibitem{peak_pic}
See Fig. 13(d) in Ref. \cite{chate_08} and Fig. 8(c) in \cite{bertin_09}.

\bibitem{romanczuk_12}
P. Romanczuk, L. Schimansky-Geier,
Ecol. Complex. {\bf 10}, 83 (2012).

\bibitem{grossmann_13}
R. Gro{\ss}mann, L. Schimansky-Geier, P. Romanczuk,
arXiv:1301.5890v2 (2013).

\bibitem{mishra_10}
S. Mishra, A. Baskaran, M.C. Marchetti,
Phys. Rev. E {\bf 81}, 061916 (2010).

\bibitem{chou_12}
Y.-L. Chou, R. Wolfe, T. Ihle,
Phys. Rev. E {\bf 86} 021120 (2012).

\bibitem{aldana_09}
M. Aldana, H. Larralde, B. Vazquez, Int. J. Mod. Phys. B {\bf 23} 3661 (2009).

\bibitem{wave_num}
See Fig. 13 in Ref. \cite{chate_08} and Fig. 8 in \cite{bertin_09}.

\bibitem{phasediag_num}
See Fig. 2(a) in Ref. \cite{chate_08} and Fig. A1(b) in \cite{bertin_09}.


\bibitem{bag_13}
G. Baglietto (2013), private communication.

\bibitem{ihle_09}
T. Ihle, Phys. Chem. Chem. Phys. {\bf 11}, 9667 (2009).

\bibitem{lattice_boltz}
X. He, L-S. Luo, Phys. Rev. E {\bf 56}, 6811 (1997);
S. Chen, G.D. Doolen, Ann. Rev. Fluid Mech. {\bf 30}, 329 (1998).

\end{thebibliography}
\end{document}